\documentclass[12pt]{iopart}
\usepackage{iopams}
\usepackage{graphicx}

\newcommand{\eqref}[1]{(\ref{#1})}

\begin{document}

\title[Frequency tuning, nonlinearities and mode coupling in circular graphene resonators]{Frequency tuning, nonlinearities and mode coupling in circular mechanical graphene resonators}

\author{M. Eriksson, D. Midtvedt,  A. Croy, A. Isacsson}

\address{Department of Applied Physics,
Chalmers University of Technology
SE-412 96, G\"{o}teborg, SWEDEN}
\ead{andreas.isacsson@chalmers.se}
\begin{abstract}
We study circular nanomechanical graphene resonators by means of
continuum elasticity theory, treating them as membranes. We derive
dynamic equations for the flexural mode amplitudes. Due to geometrical
nonlinearity these can be modeled by coupled Duffing equations. By
solving the Airy stress problem we obtain analytic expressions for
eigenfrequencies and nonlinear coefficients as functions of radius,
suspension height, initial tension, back-gate voltage and elastic
constants, which we compare with finite element simulations.  Using
perturbation theory, we show that it is necessary to include the
effects of the non-uniform stress distribution for finite
deflections. This correctly reproduces the spectrum and frequency
tuning of the resonator, including frequency crossings.
\end{abstract}

\maketitle

\section{\bf  Introduction}
In the field of nanoelectromechanical (NEM) systems, nonlinear dynamic
phenomena such as bifurcations and mode coupling are receiving
increasing interest \cite{Karabalin, Bachtold_Damp,
  Bachtold_CNT_coupl, Roukes_NL_2013, Kotthaus, Mahboob, Yamaguchi,
  Poot_Zant, Venstra_Zant,Buks}. In particular, graphene
resonators~\cite{Bachtold_CNT_coupl,Bunch, Craighead, Hakonen, Hone1,
  Hone2, Desmukh1, Desmukh2, Zettl}, are known to display strong
geometric nonlinear conservative as well as nonlinear dissipative
response~\cite{Bachtold_Damp,Hakonen, Hone1, Atalaya_NL,
  Croy_PRB}. This opens up for new device
applications~\cite{Atalaya_EPL} as well as for fundamental
research~\cite{Voje_PRB,vojeInPress}. So far, however, much of the
work on nonlinear dynamics has focused on one-dimensional systems with
beam or string like behavior. In graphene resonator devices, the
geometries have mostly been rectangular and doubly clamped. Such
resonators tend to suffer from inhomogeneous strain and ill defined
mode-shapes~\cite{Bachtold_McEuen} and can possess edge modes or
scrolled edges degrading the quality factor~\cite{Park_NL}.  To
alleviate these problems, circular graphene drum resonators, which are
edge and corner free, can be used.

\begin{figure}
\begin{center}
\includegraphics[width=10 cm]{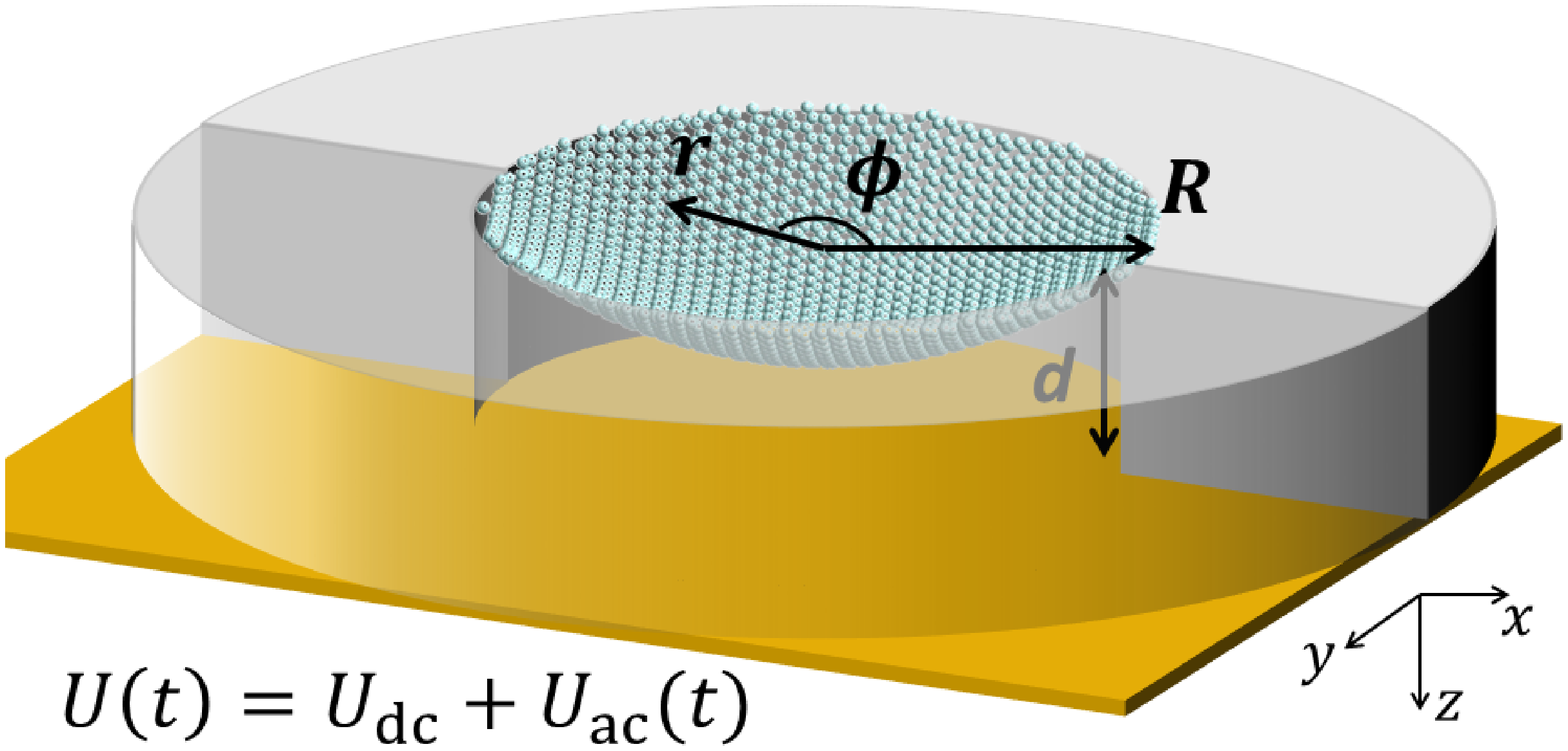}
\caption{(Color online) Circular graphene NEM-resonator of radius
  $R$. A back-gate located a distance $d$ below the membrane can tune
  and actuate the resonator by a dc-voltage $U_{\rm dc}$ and a time
  varying ac-voltage $U_{\rm ac}$, repectively.  \label{fig:sysfig}}
\end{center}
\end{figure}

Treating a circular graphene resonator (see figure~\ref{fig:sysfig})
as a membrane~\cite{Katsnelson_2013}, we derive analytic expressions
for frequency tuning, Kerr-constants and mode coupling coefficients as
functions of prestress, radius, electrostatic pressure, and elastic
parameters. Knowledge of these coefficients is required when analyzing
and designing experiments.

The study of the nonlinear dynamics of membranes and thin shells has a
long history in the field of applied mechanics and structural
engineering. For instance the static deflection solutions
corresponding to the Hencky problem~\cite{old_reference1,
  old_reference2} and fundamental mode
nonlinearity~\cite{old_reference3}, have been studied previously; for
a review see Ref.~\cite{Jenkins_1996}. In NEM-resonator experiments,
typically only a few flexural resonant modes are excited and
probed. The equations of motion for the normal mode coordinates
$\zeta_\alpha(\tau)$, which are nonlinear due to the geometric
nonlinearity, have the canonical form
\begin{equation}
\partial_{\tau}^2\zeta_{\alpha}+ \Lambda_{\alpha}
\zeta_{\alpha}+\sum_{\beta=1}^\infty\sum_{\gamma\geq \beta}^\infty
Q^{\alpha}_{\beta\gamma} \zeta_{\beta}\zeta_{\gamma}
+\sum_{\beta=1}^\infty\sum_{\gamma\geq \beta}^\infty\sum_{\eta\geq
  \gamma}^\infty C^{\alpha}_{\beta\gamma\eta}
\zeta_{\beta}\zeta_{\gamma}\zeta_{\eta} =f_\alpha(\tau),
\label{eq:gensystem}
\end{equation}
where $f_\alpha(\tau)$ represents external forcing.  This form
facilitates the description and analysis of the system dynamics. Using
first order perturbation theory, and comparing with finite elements
simulations, we give closed expressions for $\Lambda_\alpha$ and
coupling constants $Q_{\beta\gamma}^{\alpha}$ and
$C_{\beta\gamma\eta}^{\alpha}$. We find that significant contributions
to the frequency spectrum come from inclusion of the
deflection-induced non-uniform tension. In particular it gives rise to
frequency crossings, i.e., an interchange of mode frequencies with
increasing back-gate voltage.

The organization of this paper is as follows: First, in section
\ref{continSec}, we present and discuss the validity of treating a
graphene resonator as a membrane. Then, in section \ref{statSec}, we
consider the problem of finding the static shape and accompanying
non-uniform tension profile in the presence of a static back-gate
voltage. We then, in section \ref{modeExpSec}, reformulate the problem
in terms of the coupled Duffing equations~\eqref{eq:gensystem}, while
section \ref{linSec} treats the frequency spectrum of the lowest lying
modes as well as the frequency tuning with back-gate voltage. Finally,
in section \ref{nonlinSec} we calculate the quadratic and cubic
nonlinear coupling constants appearing in (\ref{eq:gensystem}).

\section{\bf Continuum description of suspended graphene}
\label{continSec}
We consider a circular mechanical resonator of radius $R$ made from a
layer of graphene with a built-in uniform tensile stress $T_0$ and
suspended a distance $d$ above a back-gate, as sketched in figure
\ref{fig:sysfig}. The external forces on the membrane stem from the
gate bias voltage $U(t)=U_{\rm dc}+U_{\rm ac}(t)$.

The equations of motion for the membrane follow from the Lagrangian
density $\mathcal{L}=\mathcal{T}-(\mathcal{V}_b+\mathcal{V}_s)$
consisting of kinetic energy density $\mathcal{T}$, bending energy
density $\mathcal{V}_b=\frac{\kappa}{2}|\Delta w|^2$ and stretching
energy density
$\mathcal{V}_s=\frac{1}{2}\sigma_{ij}\epsilon_{ij}$. Here, $\kappa$ is
the bending rigidity while the stress ($\sigma_{ij}$) and strain
($\epsilon_{ij}$) components are
\begin{eqnarray}
&\sigma_{rr}=\frac{E h }{1-\nu^2} \left(\epsilon_{rr}+\nu
\epsilon_{\phi\phi} \right), & \epsilon_{rr}=\partial_r
u_r+\frac{1}{2}(\partial_r w)^2,\label{test}
\\
&\sigma_{\phi\phi}=\frac{E h }{1-\nu^2} \left(\epsilon_{\phi\phi}+\nu
\epsilon_{rr} \right), &
\epsilon_{\phi\phi}=\frac{1}{r}\partial_{\phi}
u_{\phi}+\frac{1}{r}u_r+\frac{1}{2r^2}(\partial_{\phi}
w)^2,\label{sphiphiephiphi}
\\ 
&\sigma_{r\phi}=\frac{E h}{1+\nu}\epsilon_{r\phi}, &\hspace*{-1cm}
\epsilon_{r\phi}=\frac{1}{2}\left[\left(\partial_r-\frac{1}{r}\right)
  u_{\phi}+\frac{1}{r}\partial_{\phi}
  u_r+\frac{1}{r}(\partial_r w)(\partial_{\phi} w)
  \right].\nonumber\\
&&
\end{eqnarray}
The displacement field
$\boldsymbol{u}(r,\phi)=u_r(r,\phi)\hat{r}+u_{\phi}(r,\phi)\hat{\phi}+w(r,\phi)\hat{z}$
describes the local deviation of the graphene from its relaxed
configuration when it is free of tension.  For graphene, the elastic
modulus $E$ and the mechanical equivalent membrane thickness $h$ are
combined in the 2D modulus $E h\approx 340$~N/m and the Poisson ratio
$\nu\approx0.15$ \cite{grapheneProp}. These quantities relate to the
2D Lam\'{e} coefficients $\lambda=Eh{\nu}/(1-\nu^2)$ and
$\mu={Eh}/{2(1+\nu)}$. Stationarity of the action leads to the
F\"{o}ppl-von Karman equations of motion
\begin{eqnarray}
&&\rho_0 \ddot{u}_r-\left[\partial_r \sigma_{rr}+r^{-1}\partial_{\phi}
    \sigma_{r\phi}+r^{-1}(\sigma_{rr}-\sigma_{\phi\phi})
    \right]=0,\label{reqfirst} \\ &&\rho_0
  \ddot{u}_{\phi}-\left[\partial_r
    \sigma_{r\phi}+{2}r^{-1}\sigma_{r\phi}+r^{-1}\partial_{\phi}\sigma_{\phi\phi}
    \right]=0,\label{phieqfirst}
  \\ &&\!\!\!\!\!\!\!\!\!\!\!\!\!\!\!\!\!\!\!\!\!\!\!\!\!\!\!\!\!\!\!\!\!\!\!\rho_0
  \ddot{w}+\kappa \Delta^2 w -r^{-1}\left[\partial_r(r\sigma_{rr}
    \partial_r
    w+\sigma_{r\phi}\partial_{\phi}w)+\partial_{\phi}\left(\sigma_{r\phi}\partial_r
    w +r^{-1}\sigma_{\phi\phi}\partial_{\phi}w\right)
    \right]=P_{z}(r,\phi),\label{weqfirst}
\end{eqnarray}
with mass density $\rho_0\approx 0.75$~mg/m$^2$ \cite{grapheneProp}
and external load $P_z$. For an initial uniform strain $\epsilon_0$,
the boundary conditions
supplementing~\eqref{reqfirst}-\eqref{weqfirst} are
\begin{equation}
\label{BC}
\boldsymbol{u}(r,2\pi)=\boldsymbol{u}(r,0),\ \ u_r(R,\phi)=R\epsilon_0
,\ \ u_\phi(R,\phi)=w(R,\phi)=\partial_r w|_{r=R}=0.
\end{equation}

For the electrostatic load, we adopt the local parallel plate
approximation, i.e., $P_z(r)= -(U(t)^2/2)\partial_{\tilde{z}} C$, with
$\tilde{z}(r,\phi,t)=d-w(r,\phi,t)$ and distance dependent capacitance
$C$.  Upon dividing the vertical deflection into a static and a
time-dependent part as $w(r,\phi,t)=\bar w(r,\phi)+\delta w(r\phi,t)$,
one has to lowest order in $\delta w$
\begin{equation}
\label{localLoad}
P_z=\frac{\varepsilon_0 U^2}{2(d-w)^2} \approx \frac{\varepsilon_0
  U^2}{2(d-\bar{w})^2} \left(1+2\frac{\delta w}{d-\bar{w}}\right),
\end{equation}
where $\varepsilon_0$ is the vacuum permittivity.

Before attempting to solve the problem, we can make some considerable
simplifications. Since graphene has a very low bending rigidity,
$\kappa\approx 1.5$ eV \cite{grapheneProp,Lindahl}, the ratio between
bending and stretching terms in~\eqref{weqfirst} is very small for the
lowest lying flexural vibration modes, if a built-in tension is
present \cite{PRB}. Hence, we employ the membrane approximation where
the bending rigidity is neglected. Further, since the lowest
frequencies of in-plane vibrations are typically one or two orders of
magnitude larger than the frequencies of the lowest lying out-of-plane
vibrations, we can treat the in-plane displacements
adiabatically. Hence, we drop the terms $\ddot{u}_r=\ddot{u}_\phi=0$
in \eqref{reqfirst} and \eqref{phieqfirst} and solve the ensuing
equilibrium equations as functions of $w$.

To solve the simplified in-plane problem we introduce the Airy stress
field $\chi$ \cite{timoPlateShell} satisfying the inhomogeneous
biharmonic equation
\begin{equation}
\Delta^2 \chi(r,\phi,t) = {Eh}{R^{-2}}\ F[w,w],\label{airy}
\end{equation}
where the source term $F[w,w]$ is bilinear in its two arguments, and defined by
\begin{eqnarray}
&&{2}{R^{-2}}F[w,w']=-(\partial_{r}^2w)\left(r^{-1}\partial_r
  w'+r^{-2}\partial_{\phi}^2w'\right)\nonumber\\ &&
  -\left({r^{-1}}\partial_r
  w+{r^{-2}}\partial_{\phi}^2w\right)(\partial_{r}^2w')+2\left(\partial_r{r^{-1}}\partial_\phi
  w\right)\left(\partial_r{r^{-1}}\partial_\phi
  w'\right).\label{bilinear}
\end{eqnarray}
By decomposing the Airy function as $\chi=\sum_n e^{in\phi}\chi_n(r)$
the general inhomogeneous problem on the disk $\Delta^2\chi=F$ can be
solved by means of integration (see Appendix A).

The function $\chi$ relates linearly to the stress components as \cite{landau}
\begin{equation}
\label{stressAiry}
\sigma_{rr}=T_0 + {r^{-1}}\partial_r \chi+{r^{-2}}\partial_{\phi}^2\chi,\ \ \ 
\sigma_{\phi\phi}=T_0 + \partial_{r}^2 \chi,\ \ \ 
\sigma_{r\phi}=-\partial_r\left(r^{-1}\partial_{\phi}\chi\right).
\end{equation}
The problem is then reduced to solving the nonlinear out-of-plane equation
\begin{equation}
\label{weqfirstad}
\hspace*{-1.5cm}\rho_0 \partial_t^2{w} -{r}^{-1}\left[\sigma_{rr} r\partial_{r}^2w
  +2\sigma_{r\phi}\partial_\phi\left(\partial_{r}-r^{-1}\right)w
  +\sigma_{\phi\phi}\left(\partial_{r}+{r^{-1}}\partial_{\phi}^2
  \right)w \right]=P_{z},
\end{equation}
where we obtain $\sigma_{ij}$ from solving \eqref{airy} and using the
relations \eqref{stressAiry}.

For convenience we will from here on use dimensionless variables
$\tau=tR^{-1}\sqrt{{T}/{\rho_0}}$, $\rho={r}/{R}$,
$\upsilon_r={u_r}/{R}$, $\upsilon_\phi={u_\phi}/{R}$, $\zeta={w}/{R}$,
and $\Phi_z =P_z{R}/{T}$, where $T$ is the uniform part of the stress
including the stiffening when deflecting the membrane, defined in
\eqref{tEquation}.

\section{\bf Static deformation} 
\label{statSec}
To address the dynamics, we must solve the static problem to find the
time-independent components of the deformation field which we denote
by $(\bar \upsilon_r,\bar \upsilon_\phi, \bar \zeta)$. As the load has
axial symmetry, $\bar{\upsilon}_\phi\equiv 0$, while
$\bar{\upsilon}_r$ and $\bar{\zeta}$ are found from solving
\eqref{weqfirstad} with $\partial_t^2w=0$. To this end we make the
Ansatz $\bar{\zeta}(\rho)=\zeta_0\left(1-\rho^2\right)$ where the
scaled maximum deflection at the center of the resonator
$\zeta_0=w_0/R$ is the variational parameter. With this Ansatz one
finds for the static radial displacement
\begin{equation}
\label{urofwstat}
\bar{\upsilon}_r(\rho)=\rho\left[(T_0/Eh)(1-\nu)+(\zeta_0/4)(3-\nu)\left(1-\rho^2\right)\right],
\end{equation}
and that $\zeta_0$ is found from solving a cubic equation. The latter
gives
\begin{equation}
\label{solutionW0}
\begin{array}{c}
\zeta_0= b\left({2}/{3\theta}\right)^{1/3}-\left({ \theta }/{18
  c^3}\right)^{1/3},\ \ \ \theta=\sqrt{3}\sqrt{4 b^3 c^3+27 a^2c^4}-9
ac^2.
\end{array}
\end{equation}
The constants appearing in \eqref{solutionW0} are $a=-{ R
  \varepsilon_0 U_{\rm{dc}}^2}/{4Eh d^2}$, $b=2{T_0}/{Eh} + {2 R^2
  \varepsilon_0 U_{\rm{dc}}^2}/{3 Ehd^3}$, and $c=(7-\nu)/6(1-\nu)$.
If $\bar{\Phi}_z$ is approximated as a uniform load one finds
$b=2T_0/Eh$.

\begin{figure}\centering
\includegraphics[width=14cm]{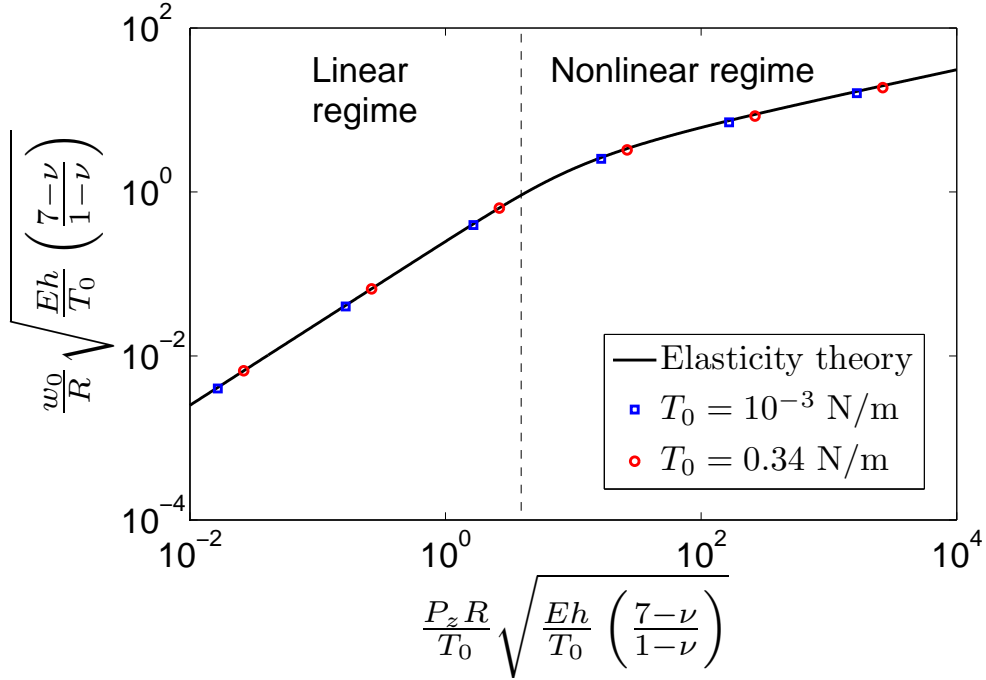}
\caption{\label{comsolStatic} (Color online) Comparison of the static
  vertical deflection at the center of the membrane $w_0=w(r=0)$, due
  to a uniform electrostatic pressure $P_z$. Symbols result from
  COMSOL Multiphysics simulations with low
  (squares) and high (circles)
  uniform prestress $T_0$. The solid line corresponds to the solution
  obtained from the variational Ansatz $w(r)=w_0(1-[r/R]^2)$ for the
  vertical deflection and neglecting bending rigidity. For small loads
  the deflection is linear in $P_z$ and then crosses over to a
  nonlinear regime where $w_0\propto P_z^{1/3}$ \cite{landau}. }
\end{figure}

The result from the uniform load approximation is shown in figure
\ref{comsolStatic} where the maximum displacement for the variational
Ansatz has been compared to finite element simulations using COMSOL
Multiphysics.  The agreement between the analytical model and the
numerical simulations under uniform load is very good, showing that
the Ansatz and disregarding the bending rigidity are valid
approximations.

\section{\bf Mode expansion and coupled Duffing equations} 
\label{modeExpSec}
Having found the expression for the static solution we turn to the
problem of setting up the equations for small vibrations around
equilibrium. Expressing these as $\delta \zeta=\zeta-\bar{\zeta}$, the
system of differential equations for the out-of-plane deflection
(see~\eqref{airy},\eqref{weqfirstad}) can be formally stated as
\begin{equation}
\partial_{\tau}^2\delta\zeta-[\hat L+\hat K]\delta
\zeta=\delta\Phi_z.
\label{opeq}
\end{equation}
Here $\delta\Phi_z$ is the time dependent part of the external force,
$\hat L$ is a linear operator and $\hat K$ an operator corresponding
to the nonlinear part of the problem. Both operators depend on the
static deflection $\bar \zeta$.

To obtain equations for the mode amplitudes, we expand the vertical
displacement in eigenmodes of the eigenvalue problem $\hat
L\Psi_\alpha +\Lambda_\alpha \Psi_\alpha=0$ as
\begin{equation}
\label{wDecomp}
\zeta(\rho,\phi,\tau)=\bar{\zeta}(\rho)+
\sum_\alpha\zeta_{\alpha}(\tau)\Psi_{\alpha}(\rho,\phi).
\end{equation}
The composite mode index $\alpha=(n_\alpha,k_\alpha)$ consists of an
angular component $n_\alpha$ and a radial component $k_\alpha$, such
that
$\Psi_\alpha(\rho,\phi)=e^{in_\alpha\phi}\Psi_{n_\alpha,k_\alpha}(\rho).$

To obtain the coefficients in Eq.~(1) it is useful to first consider
the stress-fields. As the stresses in~\eqref{stressAiry} are linear in
$\chi$ which, in turn, is bilinear in the arguments of $F$
(see~\eqref{bilinear}), also the stresses will be bilinear in the
arguments of $F$. Hence, using bilinearity we can decompose the
stress-fields corresponding to a displacement
$\zeta=\bar\zeta+\delta\zeta$ into a static part and time dependent
linear and nonlinear parts in $\delta\zeta$, i.e.,
\begin{equation}
\label{sigmaexp}
\sigma_{ij}[\zeta,\zeta]=T[\bar{\zeta},\bar{\zeta}]+\bar{\sigma}_{ij}[\bar{\zeta},\bar{\zeta}]
+2\sum_\alpha \zeta_\alpha
\sigma_{ij}[\bar{\zeta},\Psi_\alpha]+\sum_{\alpha\beta}\zeta_\alpha
\zeta_\beta\sigma_{ij}[\Psi_\alpha,\Psi_\beta].
\end{equation}
The notation $\sigma_{ij}[f,g]$ indicates that this stress is found
from solving $\Delta^2\chi=EhR^{-2}F[f,g]$.  We have further divided
the static stress into one uniform part $T$ and one spatially
dependent part $\bar{\sigma}_{ij}$.  Explicit calculations show that
\begin{equation}
\label{tEquation}
T=T_0\left(1+{x^2}/{4}\right),\ \ \ x=\zeta_0({Eh}/{T_0})^{1/2}\sqrt{(3-\nu)/(1-\nu)}.
\end{equation}

The coupled mode equations are obtained from inserting the expansion
\eqref{wDecomp} in the out-of-plane equation \eqref{weqfirstad} along
with the expansion \eqref{sigmaexp} and projecting it onto each
eigenmode $\Psi_\alpha$. This leads to the desired system of nonlinear
coupled Duffing equations
\begin{equation}
\label{duffing}
\partial_{\tau}^2\zeta_{\alpha}+ \Lambda_{\alpha}
\zeta_{\alpha}+\sum_{\beta=1}^\infty\sum_{\gamma\geq \beta}^\infty
Q^{\alpha}_{\beta\gamma} \zeta_{\beta}\zeta_{\gamma}
+\sum_{\beta=1}^\infty\sum_{\gamma\geq \beta}^\infty\sum_{\eta\geq
  \gamma}^\infty C^{\alpha}_{\beta\gamma\eta}
\zeta_{\beta}\zeta_{\gamma}\zeta_{\eta} =\langle \Psi_\alpha^* \delta
\Phi_z \rangle.
\end{equation}
Here, $\langle \Psi_\alpha^*\Psi_\beta
\rangle=\int_0^{2\pi}\int_0^1\textrm{d}\phi\textrm{d}\rho\ \rho\Psi_\alpha^*\Psi_\beta$,
where $ ^*$ denotes complex conjugation. We can now identify the
linear operator in \eqref{opeq} as $L=\nabla^2+\hat V$ with
\begin{eqnarray}
&&\hspace*{-2.2cm}\hat{V}\delta \zeta
=T^{-1}\left(\bar{\sigma}_{rr}\partial_{\rho}^2\delta \zeta
+\bar{\sigma}_{\phi\phi}\left(\rho^{-1}\partial_{\rho}+\rho^{-2}\partial_{\phi}^2
\right)\delta \zeta+2\sigma_{rr}[\bar{\zeta},\delta
  \zeta]\partial_{\rho}^2\bar{\zeta}+2\sigma_{\phi\phi}[\bar{\zeta},\delta
  \zeta]\rho^{-1}\partial_\rho \bar{\zeta}\right).\nonumber\\&&\label{Q}
\end{eqnarray} 

The overlaps with the nonlinear operator $\hat K$ give rise to the
double sum containing the constants $Q_{\beta\gamma}^\alpha$ and the
triple sum containing the constants
$C_{\beta\gamma\eta}^{\alpha}$. The quadratic coupling constants are
defined by
\begin{eqnarray}
\hspace*{-2.2cm}Q^{\alpha}_{\beta\gamma}=-{T}^{-1}\left\langle\Psi_{\alpha}^*[2\sigma_{rr}[\bar{\zeta},\Psi_{\beta}]\partial_{\rho}^2\Psi_\gamma
  +4in_\gamma\sigma_{r\phi}[\bar{\zeta},\Psi_{\beta}]
  \partial_\rho(\rho^{-1}\Psi_{\gamma})+\sigma_{rr}[\Psi_{\beta},\Psi_{\gamma}]\partial_{\rho}^2\bar{\zeta}\right.\\ \left. +2\sigma_{\phi\phi}[\bar{\zeta},\Psi_{\beta}]\left(\rho^{-1}\partial_\rho-
          {n_\gamma^2}\rho^{-2}\right)
          \Psi_{\gamma}+\sigma_{\phi\phi}[\Psi_{\beta},\Psi_{\gamma}]\rho^{-1}\partial_{\rho}
          \bar{\zeta}]\right\rangle_{\beta\gamma},\label{quadraticCoupling}
\end{eqnarray}
where the subscripts on $\langle ...\rangle_{\beta\gamma}$ indicate
that the integral sums over all unique permutations of $\beta $ and
$\gamma$. The cubic coupling constants are given by the overlaps
\begin{eqnarray}
\label{cubicCoupling}
C_{\beta\gamma\eta}^{\alpha}= -T^{-1} \left\langle\Psi_{\alpha}^*
(\sigma_{rr}[\Psi_{\beta},\Psi_{\gamma}]\partial_{\rho}^2\Psi_{\eta}+i2n_\eta\sigma_{r\phi}[\Psi_{\beta},\Psi_{\gamma}]
\partial_\rho(\rho^{-1}\Psi_{\eta})\right.\\ \left.+\sigma_{\phi\phi}[\Psi_{\beta},\Psi_{\gamma}](\rho^{-1}\partial_\rho-{n_\eta^2}\rho^{-2})\Psi_{\eta})\right\rangle_{\beta\gamma\eta}.
\end{eqnarray}

\section{\bf Frequency spectrum}\label{linSec}
The frequency spectrum of the resonator is determined by $\hat
L\Psi_\alpha+\Lambda_\alpha\Psi_\alpha$=0. As the operator $\hat V$ is
not uniform, the problem cannot, in general, be diagonalized
analytically. Instead, we will use first order perturbation theory,
treating the inhomogeneous part $\hat V$, defined in \eqref{Q} as a
perturbation. The unperturbed eigenfunctions are
\begin{equation}
\label{besselfunc}
\Psi_{n,k}^0 (\rho,\phi)={e^{i
    n\phi}J_{|n|}\left(\lambda_{|n|,k}\ \rho\right)}/{\sqrt{\pi}|J_{|n|+1}(\lambda_{|n|,k})|}
\end{equation}
for $n\in\mathbb{Z},\ k\in\mathbb{N}$. Moreover, $J_{n}(x)$ is the
$n^{\rm th}$ Bessel function and $\lambda_{n,k}$ is the $k^{\rm th}$
zero to the $n^{\rm th}$ Bessel function.  The first order correction
to the spectrum is then
$\Lambda_\alpha^{(1)}=\langle\Psi_\alpha^0\hat{V}\Psi_\alpha^0\rangle$
so that
$\Lambda_\alpha\approx\Lambda_\alpha^{(0)}+\Lambda_\alpha^{(1)}$ with
the unperturbed eigenvalues
$\Lambda_\alpha^{(0)}=\lambda_\alpha^2$. Note that the stiffening due
to the homogeneous part of the static stress is already present in the
scaling of time. Hence, the perturbation only incorporates the
inhomogeneous part of the stress.

\begin{figure}\centering
\includegraphics[width=14cm]{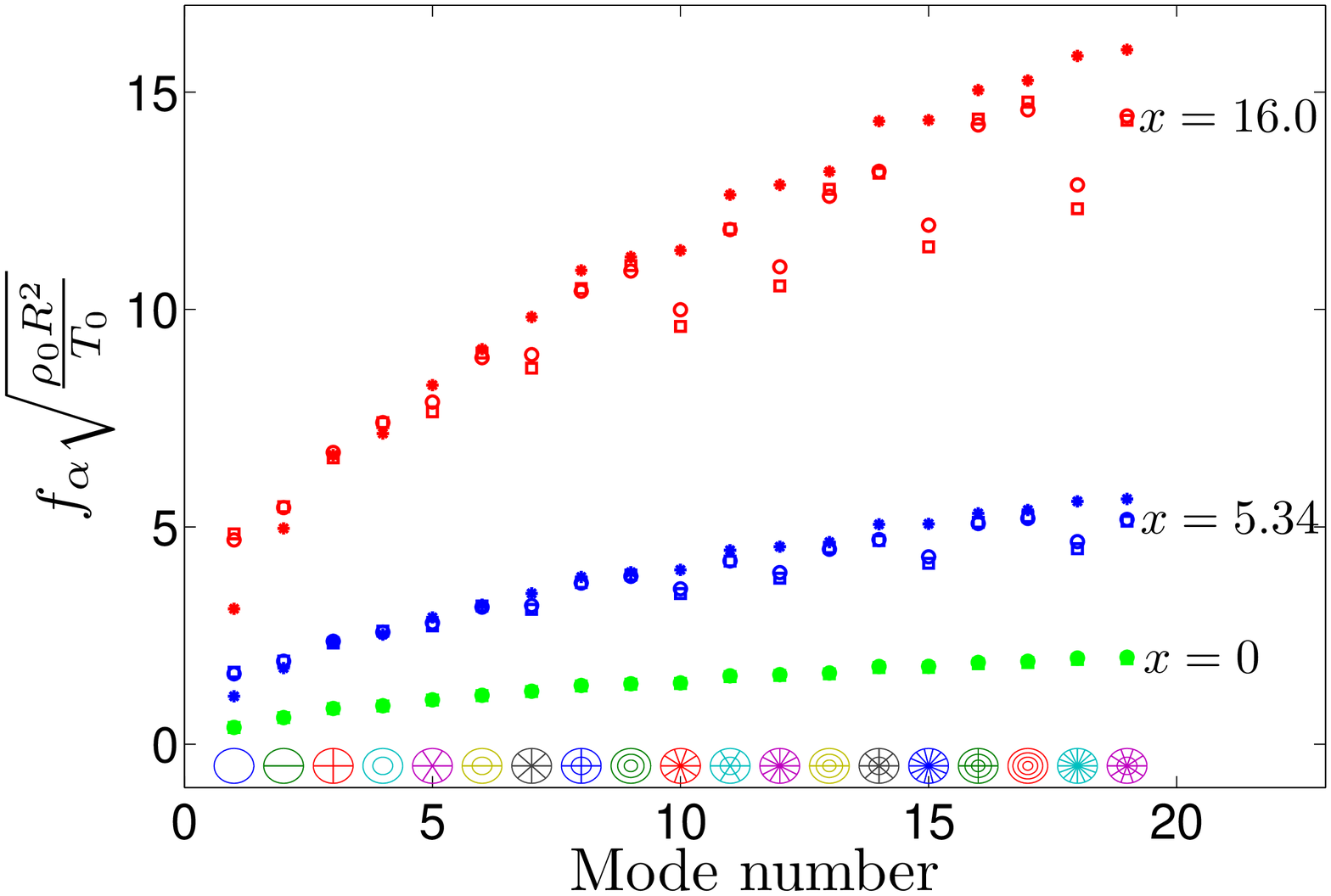}
\caption{\label{corrVsUncorr} (Color online) Frequency spectrum for
  the 19 lowest lying flexural modes for three different values of $x=
  w_0R^{-1}\sqrt{Eh({3-\nu})/T_0(1-\nu)}$. Red, blue and green symbols
  correspond to $x$ equal to 0, 5.34 and
  16.0. (Circles) Spectrum
  obtained using finite element simulations using $T_0=10^{-3}$~N/m
  and $R=1\ \mu$m. (Squares) Eigenfrequencies obtained
  using continuum theory incorporating deflection-induced non-uniform
  stress perturbatively to first
  order. (Stars) Eigenfrequencies obtained
  using only the uniform part of the deflection induced stress. Note
  that the frequency crossings are not correctly reproduced unless the
  non-uniform part of the stress is included. Here, the Poisson ratio
  $\nu=0.15$ was used.}
\end{figure}

\begin{figure}\centering
\includegraphics[width=14cm]{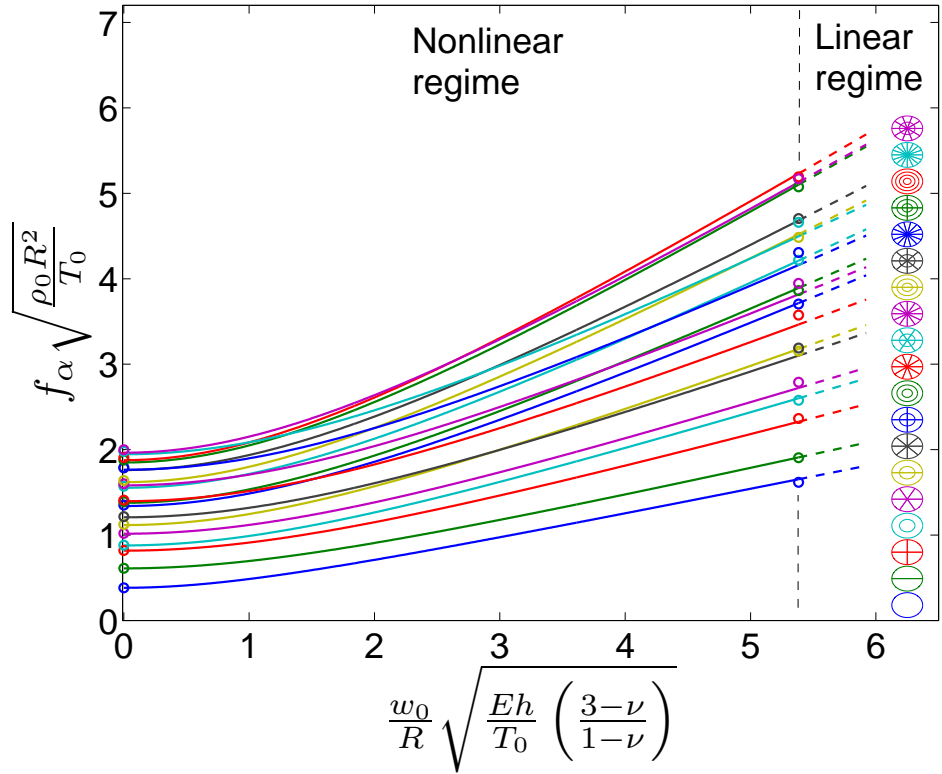}
\caption{\label{freqZoomInTest} (Color online) Frequency spectrum
  color coded to the mode shapes on the right. (Solid lines) Spectrum
  from elasticity theory including first order perturbative
  corrections due to inhomogeneous stress.  (Circles) COMSOL
  Multiphysics simulations. The majority of frequency crossings of the
  higher frequencies are taking place in the range marked ``nonlinear
  regime''. In the region marked ``linear regime'' the spectrum can be
  approximately obtained by linear extrapolation of the data in figure
  \ref{corrVsUncorr}. Here, the Poisson ratio $\nu=0.15$ was used.}
\end{figure}

Figure \ref{corrVsUncorr} shows the first order corrected frequencies
$f_\alpha=\sqrt{(\Lambda_\alpha^0+\Lambda_\alpha^1)T/(\rho_0R^2)}/(2\pi)$
under uniform load compared with the uncorrected prediction
$f_\alpha^0=\sqrt{\Lambda_\alpha^0T/(\rho_0R^2)}/(2\pi)$ along with
COMSOL Multiphysics simulations including bending rigidity. As can be
seen, the first order corrections give a good estimate of the
vibrational frequencies. A noteworthy result is that inclusion of the
first order corrections leads to a decrease in frequency for modes
with radial index $k=1$. This yields frequency crossings as the static
deflection is increased, an effect not captured by the monotonically
increasing uncorrected frequencies $f_\alpha^0$. Figure
\ref{freqZoomInTest} shows the region of parameter space where a
majority of these frequency crossings occur. For large static
deflections, $x\gtrsim5.3$, frequencies approximately increase
linearly with $x$ and can therefore be extrapolated in this region
from the data in figure \ref{corrVsUncorr}.

If the load from the back-gate is modeled by \eqref{localLoad} and
expanded to the first order in $\delta \zeta$, electrostatic softening
will shift the angular vibrational frequencies from
\begin{equation}
\label{softSimple}
\omega_\alpha^2\rightarrow\omega_\alpha^2-\omega_{\alpha,s}^2,\ \ \ \ \omega_{\alpha,s}^2\approx{\varepsilon_0U_{\rm{dc}}^2R}/{d^3\rho_0}\left(1+
H_{\alpha}{w_0}/{d}+I_{\alpha}\left({w_0}/{d}\right)^2\right).
\end{equation}
where $H_\alpha$ and $I_\alpha$ are given in table
\ref{table}. Introducing
$\omega_\alpha^0={\lambda_\alpha}{R^{-1}}\sqrt{{T_0}/{\rho_0}}$ the
expressions for the frequencies take the form
\begin{equation}
\label{freqTot}
\omega_{\alpha}=\omega_\alpha^0\sqrt{1+x^2\left(1+\Gamma_\alpha\right)/4-\left({\omega_{\alpha,s} }/{\omega_\alpha^0}\right)^2},
\end{equation}
where $\Gamma_\alpha$ is given in table \ref{table}.

\begin{figure}
\begin{center}
\includegraphics[width=14cm]{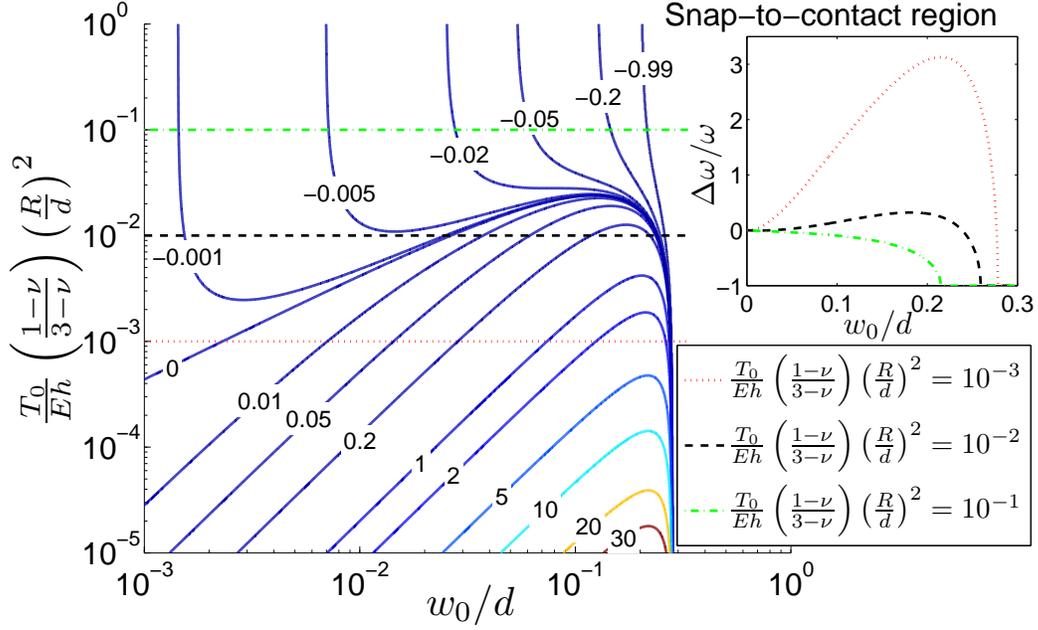}
\caption{\label{fundTuneLog} (Color online) Frequency tuning of the
  fundamental mode including electrostatic softening according to
  Eq.~\eqref{freqTot}. (Solid lines) Contours of constant relative
  frequency tuning $\Delta \omega/\omega$ of the fundamental flexural
  mode. (Inset) The frequency tuning as a function of center
  deflection $w_0/d$ for three different values of the product
  $(T_0/Eh)(R/d)^2$. The three curves correspond to the three
  horizontal lines (red dotted, black dashed and green dash-dotted) in
  the main figure. The region where snap-to-contact occurs is marked
  ``snap-to-contact region'' and takes place for deflections in the
  range $0.2<w_0/d <0.3$. Here, the Poisson ratio $\nu=0.15$ was
  used.}
\end{center}
\end{figure}

Defining $\Delta \omega_\alpha=\omega_\alpha-\omega_\alpha^0$ the
total frequency tuning, including both stiffening due to deflection
induced tension as well as electrostatic softening, can thus be
written as
\begin{equation}
\label{tuning}
\hspace*{-2.5cm}\frac{\Delta \omega_\alpha}{\omega_\alpha^0}=\sqrt{1+{x^2\frac{(1+
      \Gamma_{\alpha})}{4}}-\frac{2}{\lambda_{\alpha}^2}\left(1+
  H_{\alpha}\frac{w_0}{d}+I_{\alpha}\left(\frac{w_0}{d}\right)^2\right)\left(\frac{12+x^2\left(\frac{7-\nu}{3-\nu}\right)}{3
    -8\frac{w_0}{d}}\right)\frac{w_0}{d}}-1,
\end{equation}
where $x={w_0}{R^{-1}}\sqrt{EhT_0^{-1}(3-\nu)(1-\nu)^{-1}}$. The
tuning of the fundamental mode is plotted in figure
\ref{fundTuneLog}. For
$(Eh)^{-1}T_0(3-\nu)^{-1}(1-\nu)R^2d^{-2}>2.4\cdot 10^{-2}$ the
softening effect dominates for all deflections. With the typical
values $R=1\ \mu$m and $d=300 $ nm this corresponds to an initial
tension of $T_0=2.5$ N/m. The parallel plate approximation also
predicts a snap-to-contact region for $0.2<w_0/d<0.3$ depending on
$x$.

\section{\bf Nonlinearities and mode coupling}
\label{nonlinSec}
\begin{figure}\centering
\includegraphics[width=14cm]{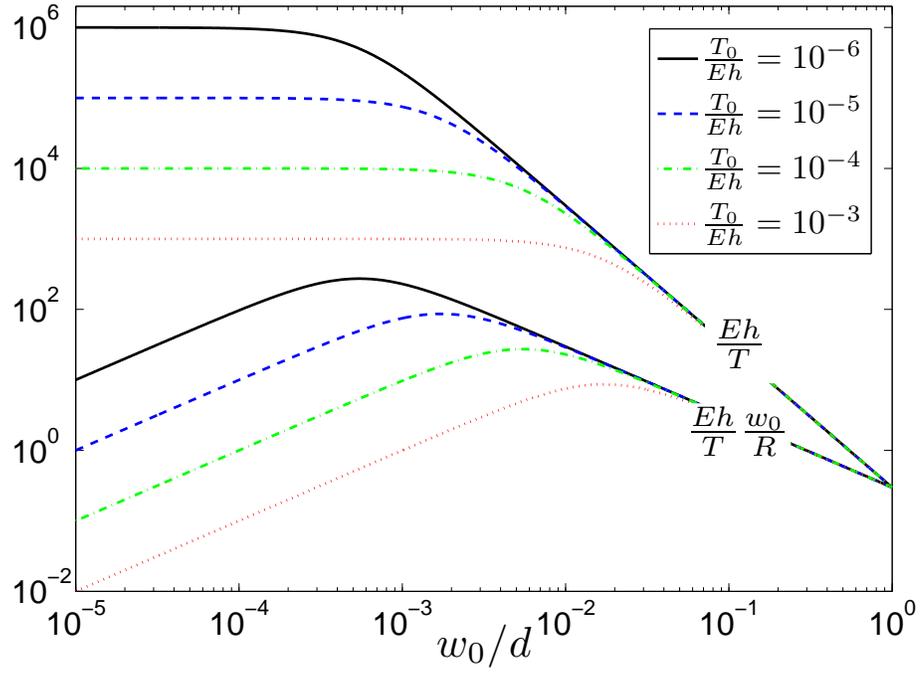}
\caption{\label{prefactors} (Color online) Prefactors of quadratic
  ($Q^\alpha_{\beta\gamma}\propto (Eh/T)[w_0/R]$) and cubic coupling
  constants ($C^\alpha_{\beta\gamma\eta}\propto Eh/T$) in
  Eq.~\eqref{duffing} as functions of center deflection $w_0/d$.}
\end{figure}

\begin{figure}\centering
\includegraphics[width=14cm]{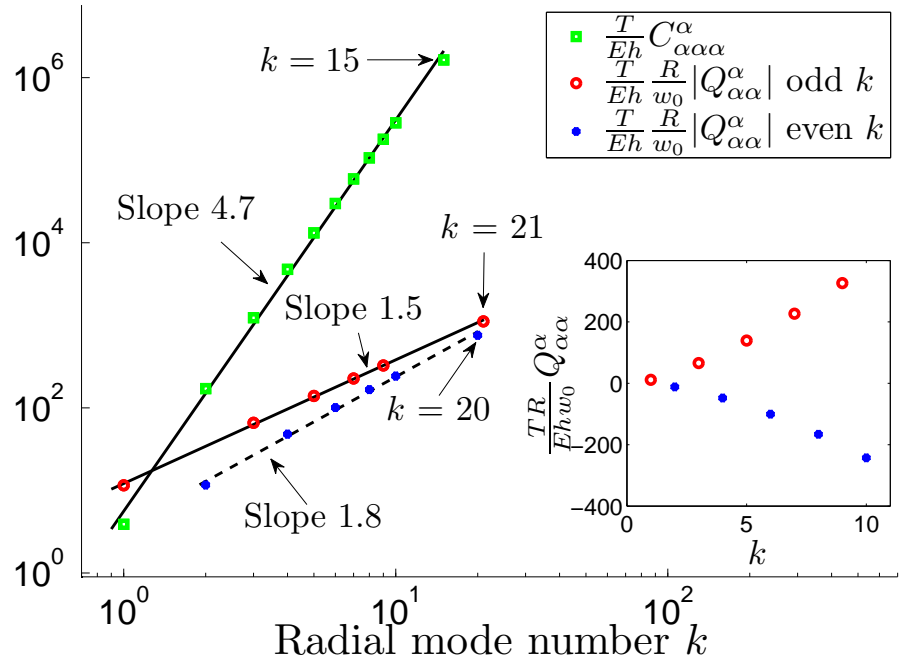}
\caption{\label{quadDiag} Diagonal quadratic and cubic coupling
  constants for symmetric modes, i.e., angular number $n=0$. Here, the
  Poisson ratio $\nu=0.15$ was used.}
\end{figure}

We finally turn to the evaluation of the coefficients of the nonlinear
terms in~\eqref{eq:gensystem}. Starting from \eqref{quadraticCoupling}
and \eqref{cubicCoupling}, we note that by angular symmetry of the
operators $\hat{L}$, $\hat{K}$ and the mode functions one finds
\begin{equation}
Q^\alpha_{\beta\gamma}\propto\delta_{n_\alpha,
  n_\beta+n_\gamma},\ \ \ C^\alpha_{\beta\gamma\eta}\propto\delta_{n_\alpha,
  n_\beta+n_\gamma+n_\eta },
\end{equation}
where $\delta $ is the Kronecker-delta. The coupling constants take
the form
\begin{equation}
\label{xIntegral}
Q^\alpha_{\beta\gamma}=\frac{Eh}{T}\frac{w_0}{R} \int_0^1 \textrm{d}
\rho\,
K^\alpha_{\beta\gamma}(\rho,\nu),\ \ \ Q^\alpha_{\beta\gamma\eta}=\frac{Eh}{T}
\int_0^1 \textrm{d} \rho\, K^\alpha_{\beta\gamma\eta}(\rho,\nu),
\end{equation}
where $K^\alpha_{\beta\gamma}$ and $K^\alpha_{\beta\gamma\eta}$
correspond to the expressions in brackets \eqref{quadraticCoupling}
and \eqref{cubicCoupling}. The prefactors are plotted in figure
\ref{prefactors}. The integrals depend on $\nu$, but the dependencies
can be extracted analytically resulting in pure numerical
integrals. By comparing with higher order corrections, we have found
that it suffices to evaluate the integrals in~\eqref{xIntegral} using
the unperturbed mode functions $\Psi^0_\alpha$ defined
in~\eqref{besselfunc}. The first order correction for the Duffing
constant of the fundamental mode reduces it with $6\%$ in the worst
case scenario. The Duffing constants $C^\alpha_{\alpha\alpha\alpha}$
for the symmetric modes are plotted in figure \ref{quadDiag} showing
power-law behavior with a small curvature correction.

The cubic and quadratic coupling constants for the 6 lowest lying
flexural modes are tabulated in tables \ref{symmCouplingTable} and
\ref{asymmCouplingTable}, respectively. They have been linearized
around $\nu=0.15$ and for small deviations from this value, $\Delta
\nu=\nu-0.15$ can to a good approximation be set to zero except for
the coupling constants $C^1_{225}$ and $C^1_{334}$.

\section{Conclusions}
Based on a continuum mechanical formulation, we have derived the
nonlinear coupled equations of motion for a circular membrane
resonator and analyzed both the eigenfrequency spectrum as well as the
nonlinear coefficients entering the equations of motion for the mode
amplitudes. For the static mode shape, due to the dc component of the
bias, we find that a simple algebraic Ansatz compares well with finite
element simulations for realistic device parameter values. For the
eigenfrequency spectrum we further find that it suffices to
incorporate the first order perturbative corrections to reproduce both
qualitatively and quantitatively the spectrum for the lowest lying
modes. In particular, we find that only by incorporating the
inhomogeneous part of the deflection-induced stress will one correctly
reproduce frequency crossings which occur with increasing static bias
voltage.  We have further derived expressions for the nonlinear
coefficients, quadratic and cubic, which must be taken into account
when modeling the dynamic response of ultrathin NEM-resonators.

\ack{The authors acknowledge funding from the European union (ME, DM,
  AI) through FP7 project no. 246026 (RODIN) and the Swedish Research
  Council VR (AC, AI).}

\newpage
\clearpage
\newpage
\clearpage
\newpage
\clearpage

\newpage
\clearpage
\newpage
\clearpage

\newpage
\clearpage

\newpage
\clearpage

\newpage
\clearpage

\begin{table}
\caption{Definitions of terms in equation \eqref{tuning} for the five
  lowest modes $\alpha=(n_\alpha,k_\alpha)$.}  \centering
\label{table}
\begin{tabular}{ l | c  c  c  l } 
    \hline\noalign{\vspace*{.7 mm}}\hline ($n$,$k$) & $H_{\alpha}$ &
    $I_{\alpha}$ &$\lambda_{\alpha}$ & $\Gamma_{\alpha}(3-\nu)$
    \\ \hline (0,1) & 2.346 & 3.859 & 2.405 & $\left({4.116 - 0.2892
      \nu}\right)$ \\ \hline (1,1) & 2.000 & 2.873 & 3.832 &
    $\left({2.046 + 0.1098 \nu }\right)
    \left(\frac{1+\nu}{3-\nu}\right)$ \\ \hline (2,1) & 1.773 & 2.290
    & 5.136 & $\left({-0.269 + 0.4577
      \nu}\right)\left(\frac{1+\nu}{3-\nu}\right)$ \\ \hline (0,2) &
    2.066 & 3.266 & 5.520 & $\left({ 0.217 - 0.0791 \nu}\right)$
    \\ \hline (3,1) & 1.607 & 1.902 & 6.380 & $\left({-1.533 + 0.7171
      \nu}\right) \left(\frac{1+\nu}{3-\nu}\right)$ \\ \hline
  \end{tabular}
  \label{freqTabular}
\end{table}
\newpage
\clearpage

\begin{table}\centering
	\caption{Cubic coupling constants between mode $\alpha$ equal
          1 to 6 corresponding to mode indices
          $\alpha=(n_\alpha,k_\alpha)$ as
          $(0,1),\ (1,1),\ (-1,1),\ (2,1),\ (-2,1)\ {\rm and}\ (0,2)$,
          respectively. By symmetry, the tables for mode 3 and 5 are
          identical to the tables for mode 2 and 4 after interchanging
          2$\leftrightarrow$3 and 4$\leftrightarrow$5. The constants
          are linearized according to $\Delta\nu=\nu-0.15$.}
	\label{symmCouplingTable}
  \begin{tabular}{ c | c  c || c | c || c | c}
 
  \hline\noalign{\vspace*{.7 mm}}\hline$\beta\gamma\eta$ &
  $\frac{T}{Eh}C^1_{\beta\gamma\eta}$ &
  $\frac{T}{Eh}C^6_{\beta\gamma\eta}$ & $\beta\gamma\eta$ &
  $\frac{T}{Eh}C^2_{\beta\gamma\eta}$ & $\beta\gamma\eta$ &
  $\frac{T}{Eh}C^4_{\beta\gamma\eta}$ \\ \hline 111 & 3.92
  +3.68$\Delta \nu$ & -1.21 & 112 & 12.5 +7.95$\Delta \nu$ & 114 &
  18.2 +17.52$\Delta \nu$\\ \hline 116 &-3.63 & 30.0 +19.4$\Delta \nu$
  & 126 & -9.13 -5.66$\Delta \nu$ & 122 & 13.2 -4.76 $\Delta \nu$
  \\ \hline 123 & 25.0+15.9$\Delta \nu$ & -6.17-2.04$\Delta \nu$ & 134
  & 29.8 -16.7$\Delta \nu$ & 146 & 63.5 -16.6$\Delta \nu$\\ \hline 145
  & 35.0+38.3$\Delta \nu$ & 91.5-24.2 $\Delta \nu$ & 223 & 69.0 +44.2
  $\Delta \nu$ & 226 & 30.8 -25.7$\Delta \nu$\\ \hline 166 &
  30.0+19.4$\Delta \nu$ & -49.0 & 245 & 121 +59.6 $\Delta \nu$ & 234 &
  121 +59.9$\Delta \nu$\\ \hline 225 & 0.509-5.57$\Delta \nu$ &
  17.6-8.39 $\Delta \nu$ & 266 & 57.5 +45.9 $\Delta \nu$ & 445 & 198
  +152$\Delta \nu$\\ \hline 236 & -12.1-9.29$\Delta \nu$ &
  115+91.8$\Delta \nu$ & 346 & 20.2 -19.4 $\Delta \nu$ & 466 &
  271+22.0 $\Delta \nu$ \\ \hline 334& 0.509-5.57$\Delta \nu$ &
  17.6-8.39$\Delta \nu$ & & & & \\ \hline 456 & 22.5+9.88$\Delta \nu$
  & 339+152$\Delta \nu$ & & & & \\ \hline 666 & -16.3 & 172
  +102$\Delta \nu$ & & & & \\ \hline
\end{tabular}
\end{table}
\newpage
\clearpage
\begin{table}\centering
	\caption{Quadratic coupling constants between mode $\alpha$
          equal 1 to 6 corresponding to mode indices
          $\alpha=(n_\alpha,k_\alpha)$ as
          $(0,1),\ (1,1),\ (-1,1),\ (2,1),\ (-2,1)\ {\rm and}\ (0,2)$,
          respectively. By symmetry, the tables for mode 3 and 5 are
          identical to the tables for mode 2 and 4 after interchanging
          2$\leftrightarrow$3 and 4$\leftrightarrow$5. The constants
          are linearized according to $\Delta\nu=\nu-0.15$.}
	\label{asymmCouplingTable}
  \begin{tabular}{ c | c  c || c | c || c | c }
   \hline\noalign{\vspace*{.7 mm}}\hline $\beta\gamma$ &
   $\frac{T}{Eh}\frac{R}{w_0}Q^1_{\beta\gamma}$ &
   $\frac{T}{Eh}\frac{R}{w_0}Q^6_{\beta\gamma}$ & $\beta\gamma$ &
   $\frac{T}{Eh}\frac{R}{w_0}Q^2_{\beta\gamma}$ & $\beta\gamma$ &
   $\frac{T}{Eh}\frac{R}{w_0}Q^4_{\beta\gamma}$ \\ \hline 11 & 11.7
   +11.3 $\Delta \nu$ & -1.32 +1.64$\Delta \nu$ & 12 &
   26.8+18.5$\Delta \nu$ & 14 & 37.8 +35.2$\Delta \nu$\\ \hline 16 &
   -2.64 +3.27 $\Delta \nu$ & 54.4 +39.6$\Delta \nu$ & 26 & 1.70
   +4.00$\Delta \nu$ & 22 & 21.4 +4.37$\Delta \nu$ \\ \hline 23 &
   28.9+21.1 $\Delta \nu$ & 6.41+9.75 $\Delta \nu$ & 34 & 18.4-2.69
   $\Delta \nu$ & 46 & 45.7 +22.3$\Delta \nu$\\ \hline 45 & 32.9 +34.5
   $\Delta \nu$ & 44.0 +22.0 $\Delta \nu$ & & & &\\ \hline 66 & 27.2
   +19.8 $\Delta \nu$ & -11.4 +25.9$\Delta \nu$ & & & &\\ \hline
      \end{tabular}
\end{table}
\appendix
\newpage
\clearpage
\section{Solving the Airy stress equation}
\label{airyAppendix}
To find the in-plane stresses we need to solve an inhomogeneous
biharmonic equation of the form $\Delta^2\chi(\rho,\phi)=f(\rho,\phi)$
on the unit disk $0<\rho <1$, $0\le \phi < 2\pi$.  Periodicity in
$\phi$ implies that we can Fourier expand $\chi$ and $f$ as
$\chi=\sum_n e^{in\phi}\chi_n(\rho)$ and $f=\sum_n
e^{in\phi}f_n(\rho)$ leaving us with the problem
$\Delta_n^2\chi_n=f_n(\rho)$, where $\Delta_n\chi_n\equiv
\rho^{|n|-1}\partial_\rho\left(\rho^{1-2|n|}\partial_\rho\rho^{|n|}\chi_n\right)$.
Repeated integration over $\rho$ shows that the general solutions can
be expressed as $\chi_n(\rho)=\chi_n^{(p)}(\rho)+H_{|n|}(\rho)$, where
$$\chi_n^{(p)}(\rho)=\rho^2\int_0^\rho {\rm d}\rho\,
G_{|n|}(\rho/\rho')\rho'f(\rho'),$$ with the Kernels
$G_0(\xi)=\frac{1}{4}\left(\left[\xi^{-2}+1\right]\ln
\xi+\xi^{-2}-1\right)$,
$G_1(\xi)=\frac{1}{16\xi}\left[\xi^2-\xi^{-2}-4\ln\xi\right]$ and
$G_{n\ge
  2}(\xi)=\frac{1}{8n}\left(\frac{1}{n+1}\left[\xi^n-\xi^{-n-2}\right]+\frac{1}{n-1}\left[\xi^{-n}-\xi^{n-2}\right]\right)$.
The terms $H_n(\rho)$ contain the nonsingular homogeneous solutions
$H_0=A_0\rho^2$, $H_1=A_1\rho^3$, and $H_{n\ge
  2}=A_n\rho^{n+2}+B_n\rho^n$. The constants $A_n$, $B_n$ are chosen
such that the in-plane displacement fields satisfy the boundary
conditions~\eqref{BC}, resulting in
\begin{eqnarray}
&&\hspace*{-2cm}A_0=-2^{-1}(1-\nu)^{-1}(\partial_\rho^2-\nu\partial_\rho)\chi_0^{(p)}|_{\rho=1},\nonumber\\ 
&&\hspace*{-2cm}A_1=-2^{-1}(3-\nu)^{-1}(\partial_\rho^2-\nu[\partial_\rho-1])\chi_1^{(p)}|_{\rho=1},\nonumber\\ &&\hspace*{-2cm}A_n=-(4n)^{-1}(3-\nu)^{-1}(2I_1(1)+(n-1)[I_2(1)+2I_3(1)]),\nonumber\\ &&\hspace*{-2cm}B_n=-\frac{[n(1+\nu)-2(1-\nu)]2I_1(1)+(n+1)[n(1+\nu)+2(1-\nu)][I_2(1)+2I_3(1)]}{4n^2(3-\nu)(1+\nu)},
  \nonumber
\end{eqnarray}
where
\begin{eqnarray}
I_1(x)=(\partial_x^2-\nu[\partial_x-n^2])\chi_n^{(p)},\ \ \
I_2(x)=\int_0^x \textrm{d}x'\ (\partial_{x'} \zeta(x'))^2,\nonumber\\
I_4=\nu\partial_x\chi_n^{(p)}-\int_0^x \textrm{d}x'(x'^{-1}\partial_{x'}-n^2x'^{-2})\chi_n^{(p)}.\nonumber
\end{eqnarray}

For the particular problem in this article, we need to solve a stress
problem on the form $\Delta^2 \chi=F[\zeta,\zeta]$ where $F$ is the
bilinear operator on the vertical displacement field $\zeta$,
$$F[\zeta,\zeta']=-\frac{1}{2}(\partial_{\rho}\zeta)\left(\rho^{-1}\partial_\rho
\zeta'+\rho^{-2}\partial_{\phi}^2\zeta'\right)-\frac{1}{2}\left(\rho^{-1}\partial_\rho
\zeta+\rho^{-2}\partial_{\phi}^2\zeta\right)(\partial_{\rho}^2\zeta')+\partial_\rho\rho^{-1}\partial_\phi
\zeta\left(\partial_\rho\rho^{-1}\partial_\phi \zeta'\right).$$
Writing $\zeta=\bar{\zeta}+\sum_\alpha \zeta_\alpha\Psi_\alpha$ and
using the bilinearity of $F$ one finds
$$\Delta^2 \chi=F\left[\bar{\zeta}+\sum_\alpha
  \zeta_\alpha\Psi_\alpha,\bar{\zeta}+\sum_\beta
  \zeta_\beta\Psi_\beta\right]=F[\bar{\zeta},\bar{\zeta}]+2\sum_\alpha
\zeta_\alpha
F[\bar\zeta,\Psi_\alpha]+\sum_{\alpha,\beta}\zeta_\alpha\zeta_\beta
F[\Psi_\alpha,\Psi_\beta].$$ Linearity of $\Delta^2$ allows us to
write the solution as $\chi=\bar{\chi}+2\sum_{\alpha}\zeta_\alpha
\chi^\alpha+\sum_{\alpha\beta}\zeta_\alpha \zeta_\beta\chi^{\alpha,
  \beta}$, where the terms satisfy individually the equations
$\Delta^2\bar\chi=F[\bar\zeta,\bar\zeta]$,
$\Delta^2\chi^{(\alpha)}=F[\bar\zeta,\Psi_\alpha]$,
$\Delta^2\chi^{(\alpha,\beta)}=F[\Psi_\alpha,\Psi_\beta]$.  Solution
by Fourier expansion and integration as above is now possible and we
find (using $\bar{\zeta}=\zeta_0(1-\rho^2)$)
\begin{eqnarray}
&&\bar\chi=\bar{\chi}_0=-\frac{1}{16}\ \zeta_0^2\ \rho^4+\bar{H}_0,\nonumber\\
&&\chi_n^\alpha=\zeta_0\rho \int_0^\rho \textrm{d}\rho' \Psi_\alpha(\rho') K_{|n|}(\rho/\rho')+H_{|n|}^{(\alpha)},\nonumber\\
&&\chi_n^{\alpha, \beta}=\rho^2\int_0^\rho \textrm{d}\rho'F_n[\Psi_\alpha(\rho'),\Psi_\beta(\rho')]\rho'G_{|n|}(\rho/\rho')+H_{|n|}^{(\alpha,\beta)}, \nonumber
\end{eqnarray}
with the new additional kernels $K_0(\xi)=\xi^{-1}\ln\xi$,
$K_{1}(\xi)=\frac{1}{2}\left[1-\xi^{-2}\right]$, and $K_{n\ge
  2}=\frac{1}{2n}\left[\xi^{n-1}-\xi^{-n-1} \right]$. The stress
components are then given by the derivatives \eqref{stressAiry}.
\end{document}